# A NEW ENERGY EFFICIENT APPROACH TOWARDS WASN ROUTING WITH MODIFIED QCS PROTOCOL

Debaditya Ghosh[1], Pritam Majumder[1], Ayan Kumar Das[2]

[1]Department of Computer Science & Engineering, CIEM, Kolkata, India
`debadityaghosh94@yahoo.com`, `pritmajumder@rocketmail.com`
[2]Department of Information Technology, CIEM, Kolkata, India
`ayandas24114057@yahoo.co.in`

*ABSTRACT*

*In today's world Wireless Ad-hoc sensor network, consists of small sensor nodes having limited resources, has a great potential to solve problems in various domain including disaster management. In this paper "QCS-protocol" is modified which was introduced in our previous paper [1] and named as "Modified QCS-protocol". This is the backbone of our Intelligent Energy Efficient Ad-hoc Sensor Network. Two other protocols "Irregular Information Transfer" & "Final Broadcast-Petrol Flow" protocol are also modified to enhance performance of the new version of QCS protocol to run the system properly and to make the network more energy efficient and perfect. The challenges in WASN are- limited node power, Ad-hoc organization of network and reliability. Most of the existing approaches addressed the problems separately, but not in a totality. This paper shows how the network can have unlimited life and all time readiness with overall stability to send information to the base station with minimum power dissipation with the help of multimode "same type" sensor nodes and type categorization of generated information. Moreover an effort is made to give some light to the implementation issues and analyzed overall performance of the network by MATLAB simulation.*

## 1. INTRODUCTION

In the present scenario of communication required in various fields of operation, Wireless network has an immense importance. There is a wide range of promising applications for these networks, such as, environment monitoring, health-care, battlefield operations, and other emergency and disastrous situations. Wireless network may be of various types like sensor network, Ad-Hoc sensor network, mobile network etc. Here in this paper Wireless Ad-Hoc sensor network [5], [6], is considered which consists of many small sensor nodes having limited power resources, several sensors, operating system, RAM, ROM, transceiver for wireless communications etc. Major characteristic issues in this type of network are no physical reliable connection (i.e. wireless) between the sensor nodes and no fixed network topology due to mobility of nodes. So it is clear from the characteristics of the WASN that it faces many additional challenges compared to other existing wired or non-mobile wireless networks; such as mobility of nodes, interference of signals propagated among nodes because of some environmental agents like wind, tide, temperature, thundering or due to other radio signals, network security etc.. But above all the most serious challenge is limited power resource in each node. In this paper a new dynamic multipath routing protocol is devised for the WASN network to overcome the above written challenges. The new version of our protocol is "Modified QCS protocol" which acts as a backbone protocol to run the proposed network properly having two sub protocols Irregular information transfer protocol (when the sensed data value exceeds the medium tolerance level) and Final Broad Cast– Petrol Flow protocol (When ultimate Devastating Situation occurred). As limited energy and stability of the network are the major challenges in WASN, this paper-work is done emphasizing on the energy efficiency by minimizing energy consumption of the nodes which will definitely add a positive effect on the stability of the established network and all time readiness of the network by placing and initializing the nodes in different modes ('Q' or 'C' mode) intelligently in our proposed network. Here we introduced signal type categorization depending upon the sensed values obtained from sensors and implemented this idea by using two bits of the IP-Packet header as two Flags( namely Flag1 & Flag2). We have done this categorization as we intend to minimize the number of signals (Information Packets) send- received by a node. So in our new version of the proposal in almost 40%-50% nodes will not send any signal at a time period T in regular situation, where as other 50%-60% nodes broadcast queries to its neighbors. We again used two different packet sizes to optimize the energy consumption during transmission and to secure necessary and essential information propagation.

We simulated our network in MATLAB and obtained results accordingly. We followed 6lowPAN [2], which is an acronym of IPv6 over low power Personal Area Network to fix up the Packet Header & Body size. The header compression mechanism standardized in RFC4944 [2] can be used to provide header compression of IPv6 packets so that such network can run over 802.15.4 radios. There is already three popular implementations [2] of 6lowPAN, i.e. (1) 6lowpancli, (2) B6lowpan,(3) Sicslowpan and the $1^{st}$ two are deployed in TinyOS-2.x and the $3^{rd}$ one is deployed in ContikiOS. Since all of them support UDP protocol, so that we used UDP packets to exchange information among nodes following new version of our "QCS" protocol.

## 2. STUDY OF EXISTING ROUTING PROTOCOLS

Routing in mobile ad hoc sensor networks [9] faces additional problems and challenges, when compared to routing in traditional wired networks with fixed infrastructure. There are several well known protocols in that have been specifically developed to cope with the limitations imposed by ad hoc networking environments. Now, routing protocols can be classified as Ad-Hoc protocols and Sensor network protocols.

### 2.1. Ad-Hoc Routing Protocols

Most of the existing routing protocols follow two different design approaches to handle the inherent Characteristics of ad hoc networks: the table-driven and the source-initiated on-demand approaches [3],[4].

### 2.1.1. Table-driven routing protocols

Table-driven ad hoc routing protocols maintain all times routing information regarding the connectivity of every node to all other nodes. It is also known as proactive routing and these protocols allow every node to have a clear and consistent of the network topology by propagating

periodic updates. Major examples of table-driven routing protocols are Destination-sequenced Distance-Vector Routing (DSDV) [3], Cluster head gateway Switch Routing (CGSR) [7] and Wireless routing Protocol (WRP).

### 2.1.2. Source - Initiated on-Demand Routing Protocols

An alternative approach is the source-initiated on-demand routing, also known as Reactive routing protocol. According to this approach, a route is created only when the source node requires a route to a specific destination. A route is acquired by the initiation of a route discovery function by the source node. Major examples of table-driven routing protocols are Ad-Hoc on-Demand Distance Vector Routing (AODV) [10], Dynamic Source Routing (DSR), Temporally Ordered Routing Algorithm (TORA).

## 2.2. Sensor Routing Protocols

Classification of sensor routing protocols can be done by protocol operation and network structure.

### 2.2.1. According to Protocol Operation

#### 2.2.1.1. Negotiation Based routing

These protocols use high-level data descriptors called "meta-data" in order to eliminate redundant data transmission through negotiations. The necessary decisions are based on available resources and local interactions.

#### 2.2.1.2. Multipath Based Routing

These protocols for fault tolerance by having at least one alternate path (from source to sink) and thus, increasing energy consumption and traffic generation. These paths are kept alive by sending periodic messages. It can again be classified in two types **A. Maximum Lifetime Routing in Wireless Sensor Networks** which is a protocol that routes data through a path whose nodes have the largest residual energy. The path is switched whenever a better path is discovered. The primary path will be used until its energy is below the energy of the backup path, thus achieving longer lifetime and B. **Hierarchical Power-Aware Routing in Sensor Networks** which enhances the reliability of WSNs by using multipath routing. It is useful for delivering data in unreliable environments. The idea is to define many paths from source to sink and send through them the same sub packets.

#### 2.2.1.3. Query Based Routing

In these protocols, the destination nodes propagate a query for data (sensing task or interest) from the node through the network. The node(s) containing this data send it back to the node that has initiated the query.

### 2.2.2. According to Network Structure

#### 2.2.2.1. Flat Network Routing

According to this routing protocol all the sensor nodes in the network have same role and functionality. It diffuses queries by using a set of information criteria to select which sensors to get the data.

#### 2.2.2.2. Hierarchical Network Routing

Also known as cluster-based routing, in these protocols, the nodes can play different roles in the network and normally the protocol includes the creation of clusters. Clustering algorithms in the literature varies in their objectives. Often the clustering objective is set in order to facilitate meeting the applications requirements. For example if the application is sensitive to data latency, intra and inter-cluster connectivity and the length of the data routing paths are usually considered as criteria for Cluster Head selection and node grouping clustering

### *2.2.2.3. Location Based Routing*

In the protocols, the nodes are addressed by their location. Distances to next neighboring nodes can be estimated by signaling strengths or by GPS receivers.

## 3. PROPOSED WORK

### 3.1. Basic Methodology of "Modified QCS"

One of the most important features of our approach to "Mobile Ad Hoc Sensor Network" - solution is "Generated Information Type" categorization. Depending on the type of information generated on a particular node (i.e. data values obtained from sensors after processing) or signal transmitted from a particular node to other node can be of three types- a) Regular information, b)Irregular Information, c)Devastating – Immediate Response Information. Our proposed network goes through mainly three stages – (1) Nodes placing, (2) Node activation and (3) Node communication. In this regard we had designed a basic "QCS protocol" [1] in previous paper and introduced new version of "QCS" in this paper. According to this protocol all the Nodes of this network are of same type (i.e. all the nodes have same H/w configurations) but all of them can behave in three different MODES so that depending upon the mode of communication they are named as – QUERY NODE, CHECKED NODE and SOURCE NODE. The nodes use IPv6 packets, actually 6lowPAN packets and we defined these IP packets in three categories, i.e. (1)Query_packet, (2)ACK_packet and (3)Source_packet according to their size, use in communication and packet header field-status.

### 3.2. Data Dictionary

**Table 1.** Variables list.

| 'Q' | Query Node |
|---|---|
| 'C' | Checked Node |
| 'S' | Source Node |
| Flag1 | This Flag is set in the Query_packet when the sensor node senses some irregular type of information through its sensor nodes. |
| Flag2 | This Flag is set in the Query_packet when the sensor node senses some devastating type of information through its sensor nodes. |
| T | This is the time period for which a C node will be in sleeping mode |
| Node_id | Unique identification of a node to indentify the node separately. |
| adj[n] | The node queried with own location & state information , make it as adjacent |

### 3.3. Sensor Node Details

Sensor nodes which are capable of sensing, processing, collecting information and communicating with other nodes participating in the network are the smallest functional units of any wireless sensor network thus the success of a network largely depends on the proper design of sensor nodes. The characteristics and functional specifications of the sensor nodes used in this proposed network are given below.

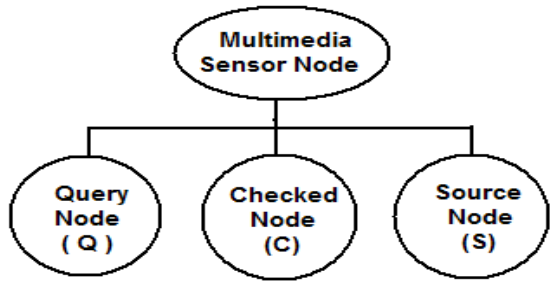 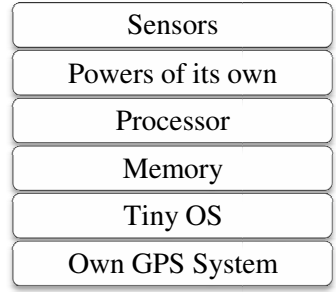

Figure1. Multimode sensor nodes   Figure2. Node components

Each sensor node in the network has same components as depicted in Figure 2. and they are of same type. Every sensor node has various sensors like temperature sensor, pressure sensor, thermo sensor, high density metal sensors etc. to collect information from the environment of its operating area. Power supply will come from rechargeable and non- rechargeable batteries, processing and decision making work is handled by a high-efficient embedded tiny OS and inbuilt memory. Nodes will also be passive GPS system enabled for the tracking purpose of a node.

But depending on the situation nodes are able to change its modes and mode wise nodes are classified into three types as shown in figure 1. Query node, Checked node and Source node.

**Query Node** Generate Query packets to its neighbors to get their status information.

**Checked Node** operates in idle mode i.e. will not query to anyone in regular network situation thus making the network energy efficient and replies to the query nodes by sending ACK packet only when Flag1 of the Query packet is set i.e. in Irregular network situation. In this mode nodes continue its processing and sensing activities irrespective network behavior.

**Source Node** is responsible for raising alarm to the network by setting its Flag(s). A node becomes source node when its Flag1 or Flag1 and Flag2 are set. Presence of source node in a particular network instance signifies that the operating area is in trouble or there is some irregularity in that domain.

The Communication media will be used for these nodes is Radio Frequency(RF) as this media require no line-of-site transmission and less sensitive to the atmospheric conditions and moreover this RF media is license free.

### 3.3. Procedure in Short

#### 3.3.1. Network Initialization Procedure

A node is activated as 'Q' node (Query node) randomly from all nodes in the network. The one non adjacent node of 'Q' is chosen and denoted as "Q". This procedure continues until all the nodes nonadjacent to each other are chosen and denoted as 'Q'. Others will be 'C' nodes, i.e. checked nodes.

### 3.3.2. Procedure to Handle Regular Information with Regular Network Behavior

This type of information propagation helps to maintain overall network stability & readiness to serve unusual /devastating events **[8]** and reduce power consumption with "Q-C-S" protocol by checking each node status like node power, temperature, pressure etc. and regular update of network condition to base station. Regular information includes current position of the nodes (longitude & latitude), node status variable's value (like node-power remained). One node queries when its status is "Q" i.e. it asks others "whether you are well or not & I am ok". At the next instance this node will remain idle (i.e. it will not send any signal) for one instance and its mode will be changed to "C". But it will do its analysis on data got from its own sensors, receive signal from Query nodes and after processing the signal it enlist the query nodes as its neighbors. This process is equally valid for all the nodes. This is how after some instances base-station will also come to know that "Network is fine". Now whenever any type of agitation occurs into the network which exceeds the upper-bound condition of being regular Flag1,Falg2 is/are set into the header of the sending signal if it is querying at that moment, otherwise checking mode, node is converted into querying node and Flag1 or/and Flag2 is/are set.

### 3.3.3. Network Behavior with Irregular Information

To maintain network integrity by node replacement and network status check and to analyze irregular events and to provide service for irregular event like unusual pressure, temperature etc. and sending information of power failure, node damage etc. to Base station by an optimized network path way.

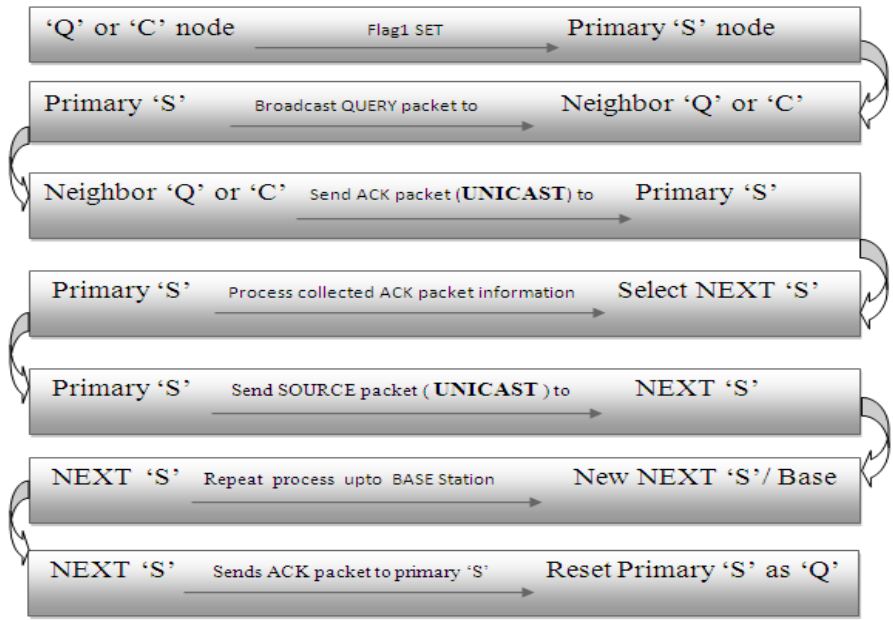

Figure 3.  Flow diagram of message passing in Irregular situation

Flag1 in the header of the signal is set, means there is some abnormal situation occurred in one or more nodes. Our intension is to track those nodes and find the shortest path to convey the information to the base station as early as possible, with minimum power dissipation, so that corrective measures can be taken quickly. Thus when Flag1 is set we switch to another algorithm

"Irregular Information Transfer*"* protocol which will handle all these things and again make all the nodes ready for devastating situations. In this irregular case path optimization to reach the base-station is not only saves time but also it saves power, because it doesn't involves all the nodes in the network.

### 3.3.3. Procedure to Handle Irregular Network behavior with Devastating Information

Information flooded to base stations following "Final Broad cast- petrol flow" network protocol. In this situation the node which will get the signal will be converted into 'S' i.e. source node and immediately it will start broadcasting that signal to others, and it will go on in a recursive manner , no node will reply to this signal .

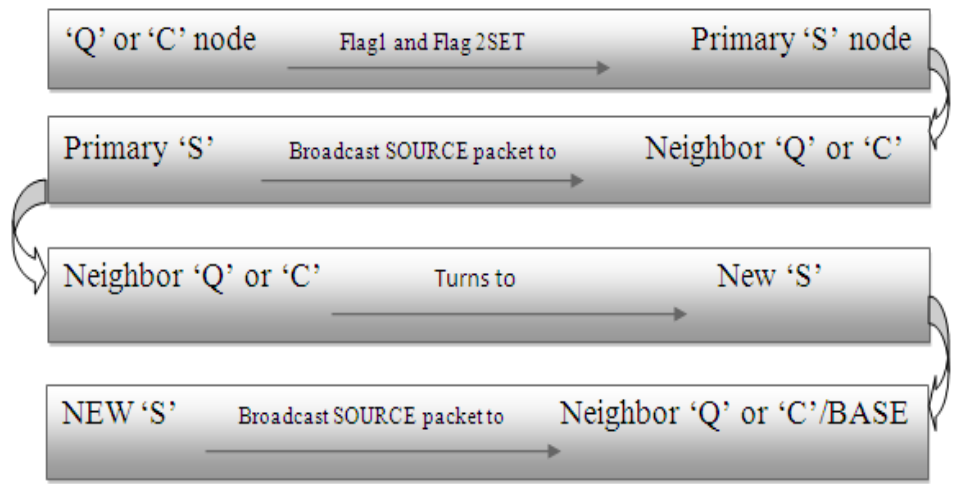

Figure 4. Flow diagram of message passing in devastating situation

## 4. DESIGN OF PROTOCOLS

"QCS" protocol is the backbone protocol of our proposed network. This protocol has two subsections – one is for network initialization and $2^{nd}$ is for after initialization of network. Two other protocols namely "Irregular Information Transfer" and "Final Broad Cast – Petrol Flow" are designed to help "QCS" protocol.

### 4.1. Algorithm for "Modified QCS" protocol

#### 4.1.1 Network Initialization

Step1. Begin.
Step2. Place the nodes and activate nodes during node placing.
Step3. Activate any one node as 'Q' node.
Step4. Activate all its neighboring nodes as 'C' node.
Step5. Activate another 'Q' node which is not a neighbor of previous 'Q' node.
Step6. Repeat Step3 to Step5 until every node on the network are activated properly.
Step7. Every node stars sensing by sensors
Step8. End.

### 4.1.2. After Initialization:

'Q' nodes send Query_packet to its neighboring nodes by broadcasting the Query_packet and 'C' nodes receives Query_packet & in both the modes data values obtained from the sensors are processed. After each time period T (timeout condition) this mode alters with each other but sensing and processing of data continues.

If at any moment Sensed value is more than the tolerable value (prefixed value) then Set node(i).mode='S' & node(i).Flag1=1 in Irregular Situation and Both the Flags are set to 1 in Devastating Situation ( irrespective of the present mode , whether its mode is 'Q' or 'C' , when next node to propagate information is chosen the previous 'S' node is reset to 'Q' or 'C' mode from where it became 'S' after getting ACK_packet from next 'S' node) . In devastating situation it will not check anything, not even the existing energy of next node.

Step1. Begin
Step2. For i=1 to n
    If (node(i).Flag1 = =0 & node(i).Flag2 = =0 & node(i).mode=='Q')
      Delay(T)
      Set node(i).mode='C'
    Else
      Set node(i).mode='Q'
    Else If (node(i).Flag1 = =1 && node(i).Flag2 = =0)
      Set node(i).mode='S'
      Go to "Irregular Information Transfer" protocol routine.
    Else If (node(i).Flag1 = =1 && node(i)Flag2 = =1)
      Set node(i).mode='S'
      Go to "Final Broadcast - Petrol Flow" protocol routine.
Step3 Go to Step2.
Step4. End.

### 4.2. Algorithm for Irregular Information Transfer protocol

Step1. Begin
Step2. node(i) Broadcast Query-Packet
Step3. neighbors check Packet Header Flag1= =1 is true then unicast ACK packet
    with present energy of that node & location( obtained from GPS)
Step4. node(i) calculates its own distance from the Base Station
    Then calculates the distances of its neighbors (i.e. nodes replied) from base station
    Then the neighbor situated nearest to base station with high energy level chosen as
      next_node, say node(j)
      Send a Source_packet to node(j) by UDP unicast
Step5.if Source_packet is received at node(j)
    then
    Set node(j).mode='S'

    Set node(j).Flag1=1
    node(j) Send unicast ACK_packet to node(i) to reset it ('RESET' written in packet
    body)
    Assign i=j ,(i.e. now the new source node is node(j) and treat it same as that of node(i))
Step7. Repeat Step2 to Step5 until Base Station receives the signal.

Step8. Action is taken by the Base Station depending on the problem.
Step9. End.

### 4.3. Algorithm for Final Broad Cast– Petrol Flow protocol

Step1. Begin.
Step2. Broadcast that "Source_packet" to the neighbors i.e. node(i).adj
Step3. Set Flag1, Flag2 of every receiving nodes and make them 'S' node.
Step4. Repeat Step2 to Step3 until base is reached && max hop-count is half of the
     number of nodes .
Step5. Signal will be received by Base Station and necessary action will be taken.
Step7. Reinitialize the network by making neighbor nodes of the Base Station and it start its
     usual job by resetting Flag1 and Falg2.
Step8. End.

## 5. NETWORK INSTANCES

Here some network instances are shown, so that the information flow among nodes can be understood clearly.

### 5.1. Regular Situation

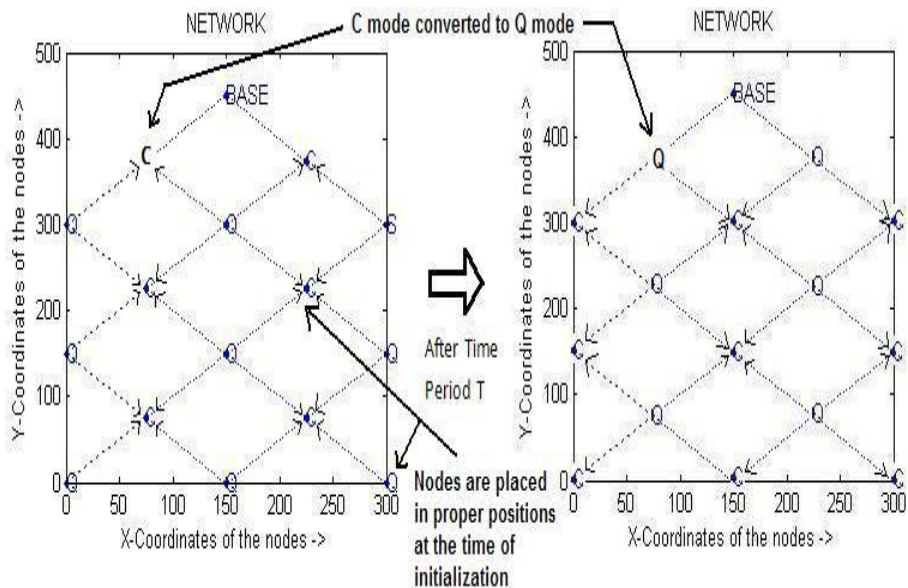

Figure 5. Regular instance

Network is initialized with 16 nodes in 300×500 test field. Nodes are placed in their predefined positions. 'Q' nodes are querying to the 'C' nodes at a time-period, say $T_i$ and at the next instant i.e. at $T_{i+1}$, 'C' nodes become 'Q' and vice versa.

## 5.2. Irregular Situation

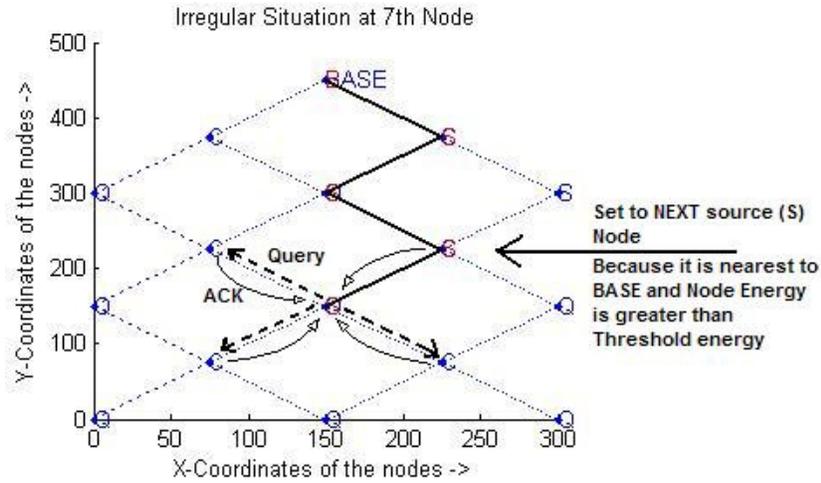

Figure 6.  Irregular instance

In the above instant at node7 at a location (150,150) some irregularity is occurred and thus this node's mode is changed to 'S' and it stars querying to its neighbors by Query_packet and the adjacent nodes replied this node with ACK_packet . Then node7 calculates distances of each neighbor and checks their energy level. Then the node with minimum distance from the base station and energy greater than threshold energy is chosen as next node and changes the mode of the next node as 'S'. This process continues until BASE, i.e. node16 is reached.

## 5.3. Devastating Situation

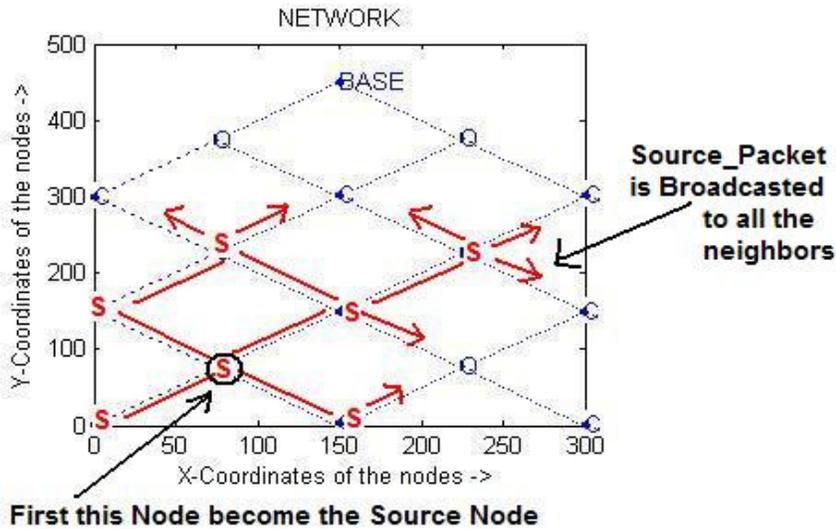

Figure 7. Devastating Instance

As shown in the above instance at node4 , at location (75,75) , extreme case occurred and thus like petrol flow the information is spreading in all direction irrespective of everything so that all the BASE stations get the information as early as possible .

90

International Journal of Ad hoc, Sensor & Ubiquitous Computing (IJASUC) Vol.2, No.3, September 2011

## 6. SOME IMPLEMENTATION ISSUES

### 6.1. Packet Structure

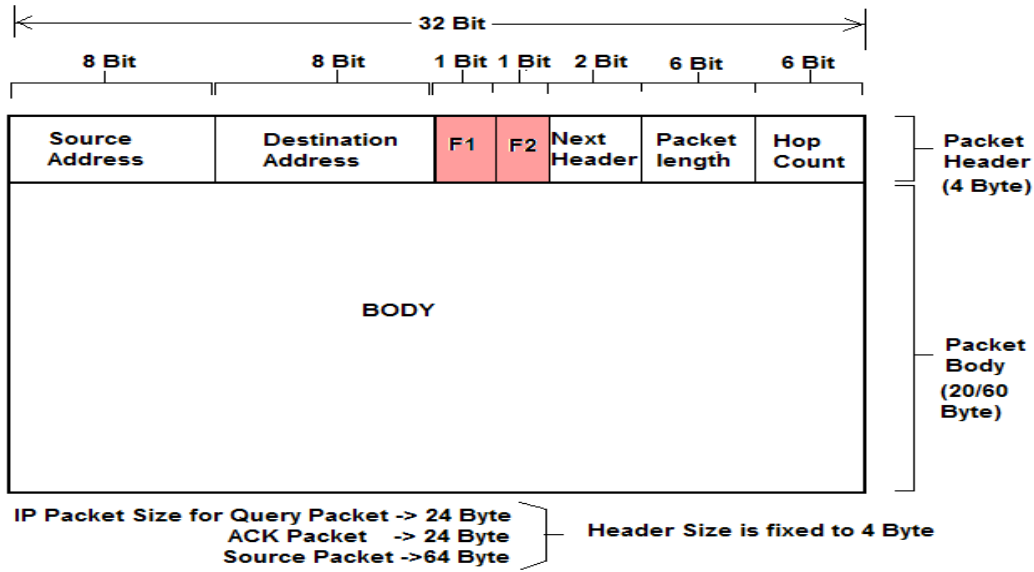

Figure 8. IP Packet Structure

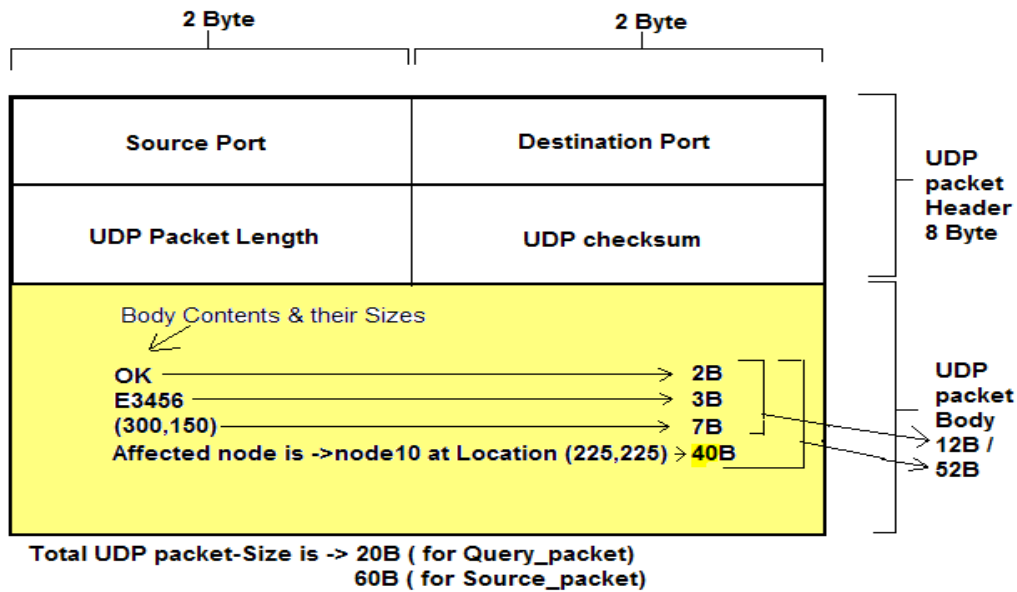

Figure 9. UDP Packet Structure

[2]Over the recent years, 6lowPAN has been proposing the use of IPv6 solutions in WSNs. IP in WSNs has been accepted not only as an extension from the conventional networks to the real world, but also as a solution to use well-known high-level protocols in WSNs, such as UDP or

HTTP. Created to adapt IPv6 long packets to IEEE802.15.4 short frames, 6lowPAN has currently three well-known implementations: 6lowpancli, b6lowpan and Sicslowpan. According to several standard compression mechanisms ( like RFC4944 , RFC4919 etc.) the IP packet Header size can be decreased from 40B (in IPv6 ) to (2-4)B to make it compatible with IEEE802.15.4 short frame formats for WASN. As per our message size we calculated that the body size in IP packet should be 20B for Query_packet and ACK_packet, 60B for Source_packet and so in total the IP packet size becomes 24B and 64B respectively. Since we all know that the UDP packet in transport layer is encapsulated into IP packet in network layer so UDP packet size is equal to the body size of IP packet.

We used just two bits in the packet header as Flag1 & Flag2 , so that in network layer when it is processed it can indicate to the signal processor to initiate the proper H/W component as the situation demands. After that, before passing to the transport layer the IP packet header will be discarded and by processing the UDP destination port number the packet body will be delivered to the correct process to analyze the message content in application layer. UDP packet body contains: (1) Node Loc, (2) Node Energy, (3) Message. In message all the required information about the source node will be provided in case of Irregular & Devastating situations so that the BASE station can take proper actions accordingly and in regular situation BASE station will come to know only that network is OK.

### 6.2. Relation between Energy and Time to send with Packet-Size [2]

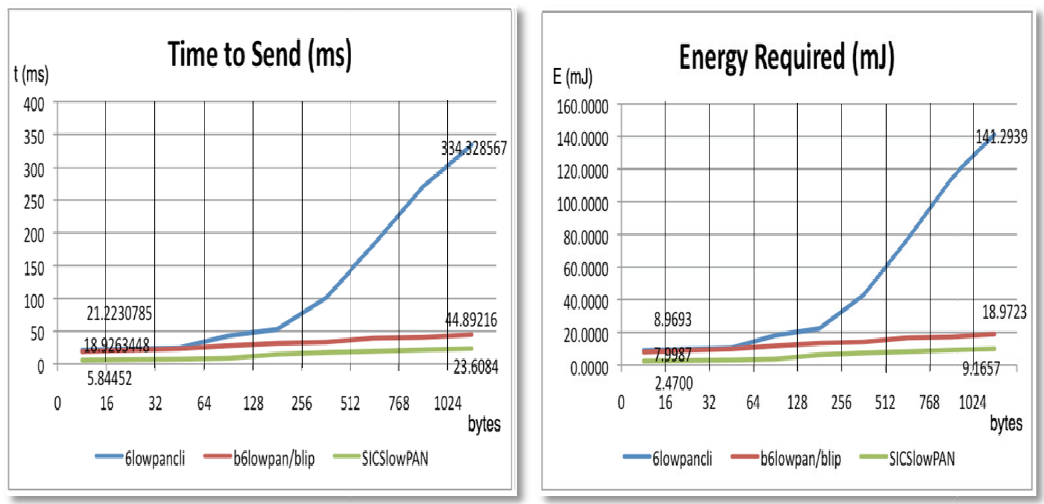

Figure 10:  Packet Size VS. Time graph [2]    Figure 11: Packet Size VS. Energy graph **[2]**

[2] Before going to the simulation results we need to fix up and clarify some more things which are very much important in this regard. Hence, considering the three implementations: 6lowpancli,b6lowpan and SICSlowPAN; and UDP packets with data length of 0, 16, 32, 64, 128, 256, 512, 768 and 1024; the time and energy required is outlined by figures 2 and 3, respectively. The time values presented in figure 2 are the average of thirty measurements, with a confident interval of 95%. It is possible to conclude that the energy required is proportional to the time spent. Considering that the telosB in tests [2] required ss18.7mA (measured using a multimeter in series) to send with a voltage of 2.6Volts, the energy in joules is given by:

E= t × 18.7 × 2.6 where *t* is the time measured.

International Journal of Ad hoc, Sensor & Ubiquitous Computing (IJASUC) Vol.2, No.3, September 2011

From the graph above incase of 6lowPAN the time required to send 24B packet is 20ms (approx.) and for 64B packet size it takes 40ms (approx.). Since Energy required is function of time as already proved we can come down to a generalized relation that Source_packet of size 64B will consume double Energy than Query_packet & ACK_packet of size 24B .

Thus for simplicity in simulation we at first assigned energy in a range 3000-5000 units randomly to each nodes and assumed that Query_packet & ACK_packet consumes 1unit and Source_packet consumes 2 units of energy. We assumed the threshold energy as 500 units under which energy value will be alarming.

## 7. SIMULATION & RESULTS

### 7.1 Simulation workspace:

As we mentioned earlier that we have done the simulation work in MATLAB tool to generate network with nodes having several properties. We generated several graphs to analyze the performance of the simulated network so as to get some idea about the behavior of the network in real world. After showing the results we directly enter into the performance analysis and explain each result obtained from simulation separately.

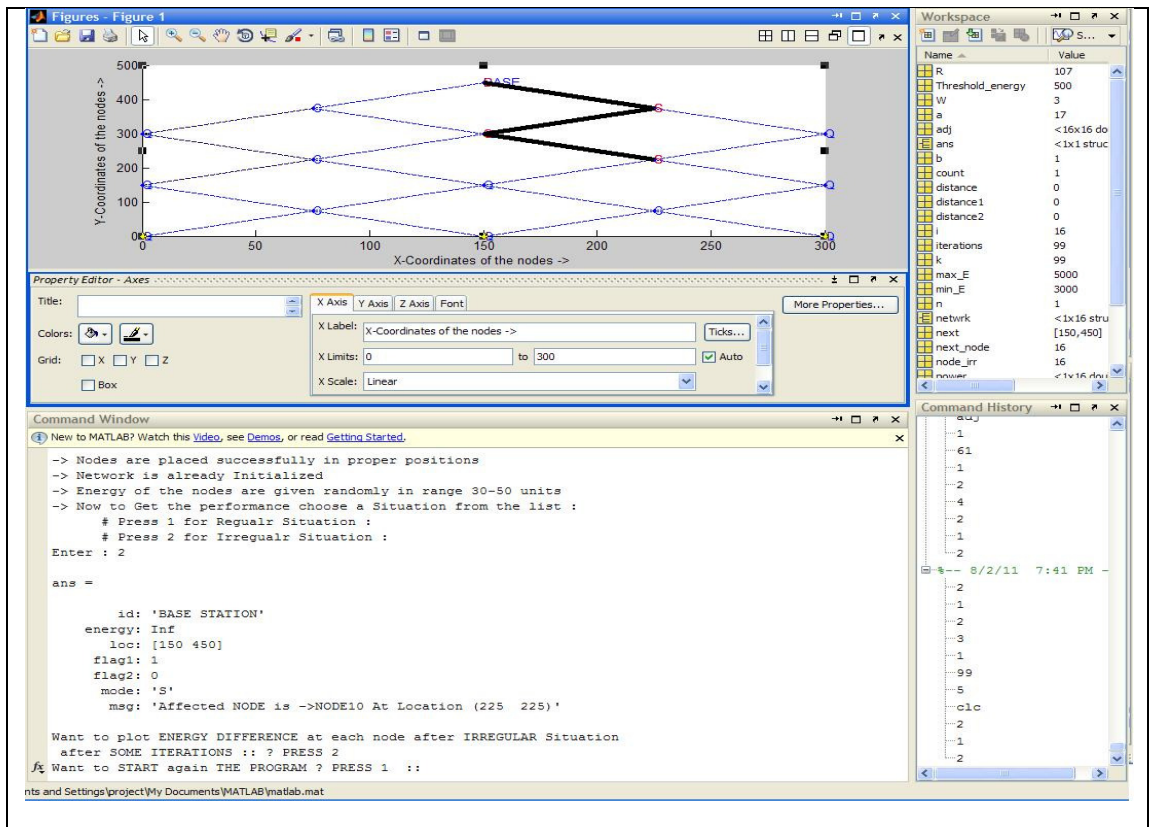

Fig 12: Workspace Instance

In the above snap-shot Irregular information generated at node10
That information is propagated to the Base Station through a optimized path
node10 – > node12 –> node15 –> node16/BASE

International Journal of Ad hoc, Sensor & Ubiquitous Computing (IJASUC) Vol.2, No.3, September 2011

The BASE STATION status in this snap is as follows:

```
ans = 

        id: 'BASE STATION'
    energy: Inf
       loc: [150 450]
     flag1: 1
     flag2: 0
      mode: 'S'
       msg: 'Affected NODE is ->NODE10 At Location (225  225)'
```

### 7.3. Results Obtained After Simulation

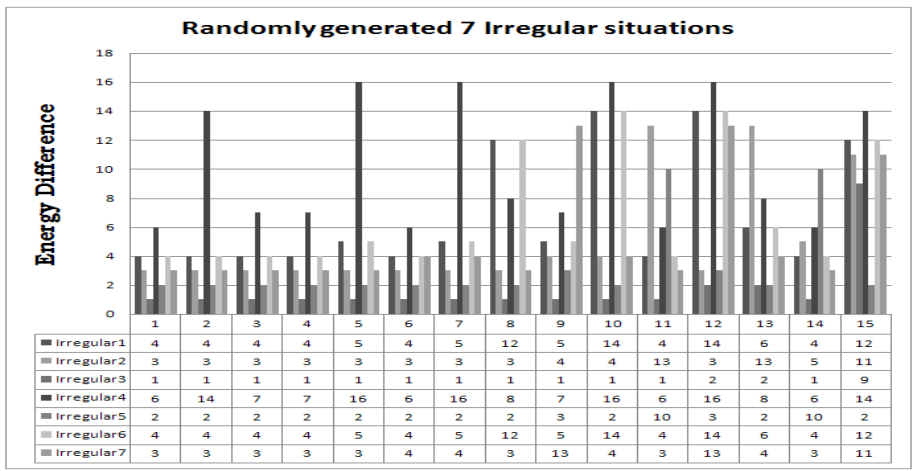

Fig 13: Energy Difference( Y-axis) VS. node_ID graph (X-axis)  Randomly selected Source nodes are N13, N12, N15, N2, N14, N8, N9 respectively. Ref .Fig 6

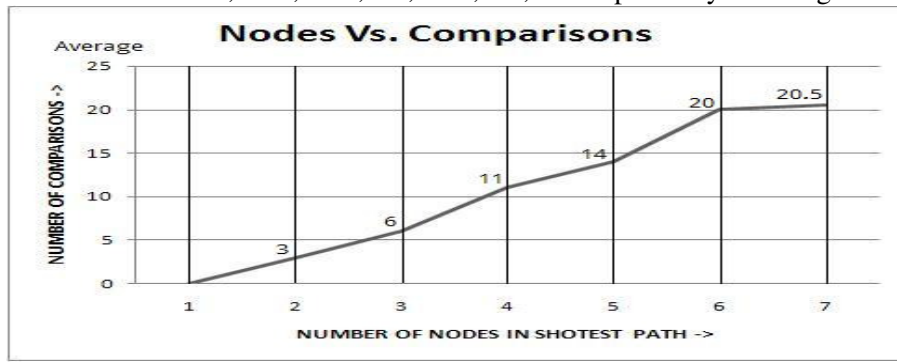

Fig 14: Number of nodes in the shortest path  formed during IRREGULAR situation VS.  number of Comparisons

94

## 8. PERFORMANCE ANALYSIS

It is the very crucial part of any paper where usually self evaluation is done supporting own proposals and may be some comparisons with others proposals are also carried out to make own stand stronger. Here we analyzed the performance of our network mainly two things keeping in mind (1) The energy efficiency of the network and (2) Life time of the network. Also we analyzed the complexity in terms of number of comparisons Fig.14 to select the optimized path required to reach information to the BASE station initially selecting a node randomly as a source node and setting its Flag1 to 1, in several Irregular situations and ultimately plotting a graph. Another graph Fig.13 is drawn choosing 7 nodes randomly as source node and plotting the difference of energy of each node after one pass. We did it in a single graph so the comparative study becomes easier. The whole performance is analyzed in two approaches; one is done mathematically to get ideal or theoretical values and another with simulations to get results which are expected to be close to the practical world implementations. Though we all know that in most cases practical values and simulated results differ to a good extent, as all the real world factors can't be incorporated into the simulation program, moreover in real world there are a lot of spontaneous small or big changes which becomes the challenges for real implementation.

To calculate the life time for a node theoretically Let, Total initial energy of a node be =E
The energy consumed (send/receive) for Query_packet/ACK_packet (24B) be = $e_1$
The energy consumed (send/receive) for Source_packet (64B) be = $e_2$

From paragraph 6.2, Fig.10 and Fig.11 we can establish a relation between $e_1$ and $e_2$ i.e. $e_2=2\times e_1$. Again Let, the processing energy required in total be $e_P$ (processing energy is assumed constant as variation in processing energy requirement is negligible in comparison to the signal energy). In regular situation each node send/receive 1 Query_packet in one time interval (T) so in one time period one node will consume $(e_1+ e_P)$ energy, so obviously the theoretical life time becomes L= $E/(e_1+ e_P)$.

Now if we try to consider the excess energy consumption in irregular situations then several problems comes; like how often the irregular situation occurs, for which reason that irregular condition comes etc. and moreover we can't even assume about that situations' occurrences as that will drive us in a totally wrong direction. Thus we tried to write the Irregular algorithm with extra care so that it consumes minimum energy and involves least number of nodes to propagate information to the BASE station. So that we can ignore that energy consumption in Irregular situation while calculating theoretical lifetime of a node. To analyze the irregular situation Fig.13 we run our simulation program 7 times and each time random energy to each node is assigned and a node within test set (node1- node15) is selected randomly as source node . Then that source node follows 4.2 algorithm to find next node to be chosen and ultimately reach base station. Here again one thing needs to be clarified that in this situation nodes which are not selected in the optimized path(Fig. 5) to reach BASE , behaves normally as regular situation. Only neighbor nodes of the nodes in the optimized path need to send one ACK_packet to the source node consuming one unit energy extra than the others. As we already mentioned that for simplicity of simulation program we assumed that 24B packets consumes 1unit and 64B packets consume 2 unit of energy each time when send or received by a node (Ref. to paragraph 6.2). From the graph Fig.13 it is very clear that if source node generated is near to the BASE it will take less number of time period (T) to reach BASE and more over most of nodes in the network behaves regularly consuming least energy. For example in Irregular3 situation where N15 was chosen as source. The reverse scenario is seen in the next case i.e. in Irregular4 where N2 selected as source (see also Fig.6).Now question may arise how these energy difference values are coming. Its very

International Journal of Ad hoc, Sensor & Ubiquitous Computing (IJASUC) Vol.2, No.3, September 2011

logical and easy to understand as these energy consumptions are solely dependent on number of packets send and receive at any mode is calculated in the program. i.e. at 'Q' mode one node consumes 1unit energy as it broadcasts 1 Query_packet and 1unit by node in 'C' mode as it receive a Query_packet . In case of Irregular situation when node mode becomes 'S', at 'S' mode a node does several things – broadcasts one Query_packet, receives ACK_packet(s) from neighbor(s) , Send Source_packet to the most suitable neighbor , receives another ACK_packet send by the next chosen node for confirmation. Thus the total energy consumed at 'S' mode is = 1+ 1×number of ACK_packet received+ 2×2 + 1 = 6 + number of neighbors. So the results obtained in Fig.12 are very obvious and expected as per our assumptions. note that processing energy is included within these unit energies.

Here (Fig.14) we compromised the energy consumption of overall network with complexity of choosing next node as source node, for this reason the number of comparisons is relatively high than the number of nodes in the optimized path. For example, from the graph (Fig.13) if we consider 5 nodes (including BASE) in the path to the BASE Station then the number of comparisons are 14, i.e. almost 3 times. Again this number of comparisons will increase if topology of the network changes from our test case and number of neighbors increases. But that is also acceptable because least number of nodes are selected at the cost of high comparison so that overall network energy consumptions is minimum.

These all things i.e. energy savings are done mainly one thing keeping in mind , that is , network should be ready to face adverse situation at any time and in that situation the information to be reached as early as possible . So that we devised the algorithm which we named as "Petrol Flow Algorithm"; that will flow information to the BASE as fast as fire spread in petrol without considering anything, means in that situation when both Flag1 & Flag2 are set in a node than it will start broadcasting Source_packet and any node accepting that Source_packet will become source ('S') and start behaving like previous node. Since here only one way information propagation is used, i.e. no provision of acknowledgement it certainly reduces time consumption and since there is a possibility that some packets may be dropped, so that to serve our purpose properly all the source(s) keep broadcasting that Source_packet in a time interval (T – time period) until BASE station reset them. This process consumes huge energy in comparison to previous situations (in Regular & Irregular situation). But since it is the ultimate devastating situation, so flow information to BASE within least time is our main motto consuming any amount of energy.

## 9. CONCLUSIONS

From the above discussion it is very much clear that our simulation results came as we were expecting. This paper was written focusing a basic idea that we have to save the energy consumption by a node in most of the life time period of the node (Regular Situations), so that it can hold at least threshold energy for the most rare undesired calamity and remains ready for prompt service always. We think that idea is projected successfully through our "QCS protocol" which we modified in this paper and named as "Modified QCS-protocol". We have done a lot of modifications on "QCS-protocol" which resulted into a better and more desirable output. In regular situation case performance enhanced 30%-40%. We ensured the network stability and all time connectivity of the nodes to BASE station, so that BASE can monitor the whole network at every moment properly. We also thought about some undesirable situations, like – if one node residing at the extremity of the network, somehow going far away from the range of signal, that node itself will broadcast a signal for others when its neighbors becomes zero at any time period T consuming extra energy so that this signal can reach to other nodes, after that the BASE station

96

will come to know that node number 'X' became disconnected. At the 1st portion of the paper a brief review work is done to distinguish our proposal from the others. Our proposal is neither proactive nor reactive, basically the good characteristics of both of them are tried to be incorporated into our proposal. Again we dared to use the most recently devised 6lowPAN implementation in WASN. We also modified two bits of the IP packet header so that processing complexity and time consumption decreases and Irregular or Devastating situation can be detected at network layer itself and proper H/w can be set accordingly. Packet size is optimized in this paper. At the end of this paper simulation results accompanied by detailed performance analysis added some extra value to this paper. There are many things remained untouched also in this paper. We are working on those issues, mainly security, clustering ideas and trying to find some way so that Artificial Intelligence can be incorporated into the network. Thus we can happily conclude that our proposal will definitely help to enhance WASN performance if it is implemented in practical field.

**Authors**

1. **Debaditya Ghsoh** is a Computer Engineer and completed B. Tech from Calcutta Institute of Engineering& Management in the current year. Presently he is working on Mobile Wireless Ad-Hoc Sensor Network Design.

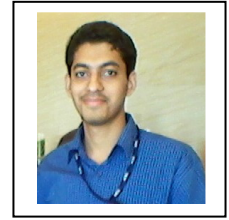

2. **Pritam Majumder** passed B.Tech this year from Calcutta Institute of Engineering and Management. Student of Computer Science. Presently working on Mobile Wireless Ad-Hoc Sensor Network Design.

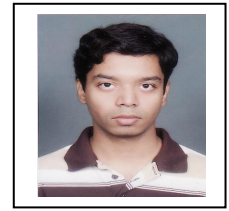

3. **Ayan Kumar Das** is an Assistant Professor in Information Technology department at Calcutta Institute of Engineering and Management. He has done M. Tech from West Bengal University of Technology and Recently doing research work on Wireless Sensor Network.

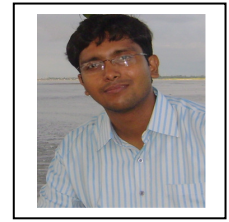